\documentclass[12pt,a4paper]{article}
\usepackage[utf8]{inputenc}
\usepackage{amsmath}
\usepackage{amsfonts}
\usepackage{amssymb}
\usepackage{graphicx}
\usepackage{xcolor}
\numberwithin{equation}{section}
\usepackage{srcltx}
\numberwithin{equation}{section}
\usepackage{authblk}
\usepackage{cancel}
\usepackage[left=2.00cm, right=2.00cm, top=2.00cm, bottom=2.00cm]{geometry}

\begin{document}

\title{Gauge Symmetry Origin of  B\"acklund Transformations for  Painlev\'e Equations } 
\author[1]{V.C.C. Alves}
\author[2]{ H. Aratyn}
\author[1]{J.F. Gomes} 
\author[1]{A.H. Zimerman}
 \affil[1]{
 Instituto de F\'{\i}sica Te\'{o}rica-UNESP,
 Rua Dr Bento Teobaldo Ferraz 271, Bloco II,
 01140-070 S\~{a}o Paulo, Brazil}
\affil[2]{
Department of Physics, 
 University of Illinois at Chicago,
 845 W. Taylor St.
 Chicago, Illinois 60607-7059}

\maketitle

\abstract{
 We identify the self-similarity limit of
the second flow of $sl(N)$ mKdV
hierarchy with the periodic dressing chain thus establishing %
a connection to $A^{(1)}_{N-1}$ invariant
 Painlev\'e equations. The $A^{(1)}_{N-1}$ B\"acklund symmetries of 
dressing equations and 
Painlev\'e equations  are obtained in the self-similarity limit of gauge transformations of the mKdV
hierarchy realized as  zero-curvature  equations  
on the loop algebra $\widehat{sl}(N)$ endowed with a principal
gradation.
}	

\section{Introduction}
\label{section:intro}

We study the integrable mKdV hierarchy on the loop algebra of 
$ sl(N)=A_{N-1}$ endowed with a principal
gradation.  We show that the  second $t=t_2$ flow 
of this hierarchy  realized as 
a zero-curvature equation  becomes in the self-similarity limit the  dressing
chain invariant under the extended affine Weyl group $A^{(1)}_{N-1}$
 that gives rise to  
$A^{(1)}_{N-1}$ invariant 
Painlev\'e equations 
for all $N$ after imposing the periodicity condition.
For $N=3$ and $N=4$ cases the corresponding Painlev\'e equations 
are well-known as Painlev\'e IV and V equations and 
their relations to the self-similarity limit of the %
$sl(N)$ 
mKdV equation  was established in 
 \cite{schiff93}.
 
In this paper we present an  explicit and straightforward construction 
 of the  dressing chain and Painlev\'e equations %
in terms of mKdV objects in the self-similarity limit.
One advantage of the zero-curvature approach is that it naturally
allows for formulation of the  B\"acklund symmetries as gauge
transformations that preserve %
the algebraic structure of the gauge potentials,  see \cite{globo} and references therein. Here we utilize gauge symmetries 
of the zero-curvature equations defined on the loop algebra of $ sl(N)$
to derive the B\"acklund symmetries  of the extended affine 
Weyl group $A^{(1)}_{N-1}$ in  the context of Painlev\'e equations including higher-order Painlev\'e equations that follow from the associated integrable structure.

In order to obtain our results we connect
several observations made by various researchers 
coincidentally around the same time. In 1993, the periodic dressing chain and its
Hamilton formalism appeared as the  main focus of study \cite{Veselov-Shabat}
in the context of  the spectral theory of the Schr\"odinger
operators. Its symmetry group was shown in \cite{adler93} to be isomorphic to 
the extended affine Weyl group $A^{(1)}_{N-1}$ 
thus clearing the way to establishing  equivalence of periodic dressing chains
\cite{takasaki03,WH03,Tsuda04,SHC06} with the corresponding  Painlev\'e systems invariant under the same $A^{(1)}_{N-1}$ symmetry groups \cite{alred}.
There were other contemporary observations that now can be understood as related to these breakthroughs and  remarkably 
made in the same
year. In \cite{schiff93} the first non-trivial equation  of $sl(N)$ mKdV model 
 was  
given and shown to give rise to  Painlev\'e equations for $N=3,4$ in
the self-similarity limit. In the very same year some of the authors of
this paper pointed out the existence of discrete 
B\"acklund transformations leading to Hirota type equations for the Toda
chain of tau-functions \cite{Aratyn:1993zi}. Generalization of this
construction \cite{anpz} to generalized Volterra lattice of symmetry
transformation gave rise to an alternative  construction of 
$A^{(1)}_N$   invariant Painlev\'e systems \cite{higherpainleve} defined for
integrable systems with a different gradation than the principal 
gradation used throughout this paper.

Here we are able to explicitly tie  several of these 
observations into an unified approach by
realizing the self-similarity limit of the second flow of $sl(N)$ mKdV hierarchy
as $N$-periodic dressing chain with its parameter $\alpha=\sum_{i=1}^N
\alpha_i$ taking
values $N/2$ or zero, and derive the extended affine Weyl group $A^{(1)}_{N-1}$ 
from gauge symmetry of the second flow in the self-similarity
limit. 

The paper is organized as follows. In Section \ref{section:curvature}
we present the zero curvature approach to the $sl(N)$ mKdV hierarchy and use it to explicitly derive its  second flow that is identified in Section \ref{section:scaling} with the dressing chain and  $A^{(1)}_{N-1}$ invariant Painlev\'e equations  in the self-similarity limit. In Section \ref{section:gaugetrans} we formulate the B\"acklund symmetries of  hierarchy equations as a class of gauge transformations that maintain the matrix form of the potentials  $A_\mu, \mu=x,t$ 
that enter the zero-curvature equations  
of the integrable hierarchy. In Subsection \ref{subsection:simlimgtrans} these gauge transformations are shown to give rise to the  extended affine Weyl symmetry of Painlev\'e equations. The class of    gauge symmetry generators we are working with is such that each such generator can be factorized into  a product of maximum $N-1$  matrices depending on only one single parameter. Such one-parameter gauge  matrix can in turn  be identified  with a generator of the  extended affine Weyl group $A^{(1)}_{N-1}$. As discussed in the next  section \ref{section:highertimes} such construction further extends explicit  invariance 
under $A^{(1)}_{N-1}$ obtained from  B\"acklund gauge symmetries  
to all 
 higher-order Painlev\'e equations \cite{cosgrove} that 
appear in the self-similarity limits of  higher flows of the mKdV integrable hierarchy. These flows are conveniently labeled as $t_{n N+k}$ with $k=1,...,N-1$ and $n=0,1,...$ (see formula \eqref{a4} for an underlying graded structure behing this labeling).
We illustrate this invariance  for the mKdV hierarchy with 
 flows $t_{2 n+1}$ for $N=2, k=1$ and $n=1,2,3$.  Explicitly we will work  with  $t_3, t_5$ and $t_7$ flows of the mKdV hierarchy and the associated  Painlev\'e equations of order 
 order $2,4$ and $6$ that are all invariant under the same $A^{(1)}_{1}$ Weyl group symmetry  as  consequence of invariance of zero curvature representation under B\"acklund gauge transformation.
 
In  Section \ref{section:outlook}, that provides an outlook, we discuss  the modification of the dressing chain that maintains the Kovalevskaya-Painlev\'e property together with the part of its B\"acklund symmetry. This relates to  the program that we have pursued in several papers \cite{coalescence,p3-p5s,p3-p5} attempting to 
connect integrability to the  remaining symmetry of Painlev\'e models that survives   the partial breaking of  the  extended affine Weyl group $A^{(1)}_{N-1}$  by deformation terms that preserve some notion of integrability.

Some background material on the $sl(N)$ algebra and technical details of deriving the Painlev\'e equations out of the dressing chain are given in the two appendices to ensure the paper being self-contained.

\section{Zero-Curvature approach to mKdV equations}
\label{section:curvature}

We will be working with a loop algebra of $sl(N)$ endowed
with a principal gradation as defined in \eqref{gradoperator} 
of Appendix \ref{section:roots}, where we provide  brief account of 
$sl(N)$ Lie algebra, its roots, fundamental weights, and its 
loop algebra $ \widehat{sl}(N)$
generalization, that all constitute  algebraic foundation 
of %
zero-curvature formulation of the $sl(N)$ mKdV flows.
The starting point of the zero-curvature construction 
is the semi-simple $\widehat{sl}(N)$ element  of grade one
\begin{equation}
E^{(1)} = \sum_{k=1}^{N-1} E_{\alpha_k} + \lambda
E_{-\alpha_1-{\ldots} -\alpha_{N-1}} \, .
\label{Eone}
\end{equation}
We refer the reader to Appendix \ref{section:roots} for definitions
of  basic matrices $E_{\alpha_a}, {\ldots} , E_{\alpha_a+{\ldots} +\alpha_b}, \; 
a,b = 1,{\ldots} , N-1$.
The $t_m$ flow of 
$sl(N)$ mKdV hierarchy  is given by solving 
 the  following zero-curvature equation:
\begin{equation}
 [ \partial_x +E^{(1)} +A_0\,,\, \partial_{t_m} + D^{(0)}+ D^{(1)}+\cdots + D^{(m)} ]=0 \, ,
\label{zerocurva}
\end{equation}
where $D^{(i)} \in {\cal G}_i$ (see Appendix \ref{section:roots} for
definition of ${\cal G}_i $) are to be solved for.  Further
\[
A_0= \sum_{k=1}^{N-1} v_k h_{\alpha_k} = \left( \begin{matrix} J_1&0&\cdots
&0 \\0&J_2 &\cdots&0\\
 \vdots&\vdots &\ddots& \vdots\\ 
 0&0 &\cdots &J_{N} \end{matrix}
\right) \, ,
\]
is a diagonal element of grade zero that satisfies the trace zero condition 
$\sum_{k=1}^N J_k=0$.   The elements $h_{\alpha_k}$ of the Cartan subalgebra  are
defined in Appendix \ref{section:roots}. The following relations:
\begin{equation}
v_i-v_{i-1}=J_i, \;\;i=1,{\ldots} , N-1, \;\;\; v_0=0,\, v_N=0\, ,
\label{vivi1Ji}
\end{equation}
hold between $v_k, k=1,
{\ldots} , N-1$ and $J_i, i=1, {\ldots} ,N$. The relation $v_i =\sum_{j=1}^i J_j$
follows from \eqref{vivi1Ji} and is highly reminiscent of relation \eqref{fundweights}
between fundamental weights $\Lambda^i$ and $e_j$ vectors.

With $ \widehat{sl}(N)$ being endowed with the grading 
structure \eqref{gradingstructure} the zero-curvature equation \eqref{zerocurva} 
can be decomposed into separate equations according to their grade. 
The highest grade component of equation \eqref{zerocurva}, $\lbrack E^{(1)}, D^{(m)}
\rbrack=0$, is solved by the grade $m$ element $D^{(m)} $ in  the kernel of $E^{(1)}$. 
From now on we will focus on  the case of $m=2$ for which it follows that,
\begin{equation}
D^{(2)}=  \sum_{k=1}^{N-2} E_{\alpha_k+\alpha_{k+1}}
+ \lambda E_{-(\alpha_1+\ldots+\alpha_{N-2})}+
\lambda E_{-(\alpha_2+\ldots+\alpha_{N-1})}\, .
\label{D2}
\end{equation}
 The lower grade matrices $D^{(i)}, i=0,1$ can be solved recursively
from the appropriate grade projections of the zero-curvature equation 
\eqref{zerocurva}. For example,  the grade one element,
\begin{equation}
 D^{(1)}= \sum_{k=1}^{N-1} F_k\,  E_{\alpha_k} + F_N\, \lambda
 E_{-(\alpha_1+{\ldots} +\alpha_{N-1})}\, ,
 \label{Done}
 \end{equation}
has coefficients $F_N, F_i, i=1,{\ldots} , N-1$  :
\begin{equation} \label{Fkdefs}
F_k= v_{k+1}-v_{k-1}=J_k+J_{k+1} , \quad k=1, {\ldots} , N-1,\;\quad F_N=
v_1-v_{N-1}=J_N+J_1,\; 
\end{equation}
as determined by the grade $2$ component of the zero-curvature
 equation \eqref{zerocurva}:
\[
\lbrack  E^{(1)}, D^{(1)} \rbrack + \lbrack A_0, 
D^{(2)}\rbrack  +\partial_x D^{(2)} =0 \, .
\]
The grade $1$ projection of the zero-curvature equation \eqref{zerocurva}:
\[
\lbrack E^{(1)}, D^{(0)} \rbrack  + \lbrack A_0, D^{(1)}\rbrack  
+\partial_x D^{(1)} =0,
\]
with $D^{(0)}= \sum_{a=1}^{N-1} d_a h_{\alpha_a} $, can be cast as a vector %
equation
\begin{equation}
\partial_x {\mathbf F}  + {\mathbf V}\, {\mathbf F} = {\mathbf K}
{\mathbf d},  \qquad 
\label{frank1}
\end{equation}
with %
\[ {\mathbf F} = \left( \begin{matrix} F_1 \\ \vdots\\ F_{N-1} \end{matrix}
\right) , \quad
{\mathbf d} = \left( \begin{matrix} d_1 \\ \vdots\\ d_{N-1} \end{matrix}
\right) , %
\quad  {\mathbf v}
= \left( \begin{matrix} v_1 \\ \vdots\\ v_{N-1} \end{matrix} \right) \, ,
\]
representing key  objects as $N-1$ vectors.
In addition  ${\mathbf V}$ denotes the  diagonal matrix
\begin{equation}
{\mathbf V}_{km}=  (2 v_k-v_{k-1}-v_{k+1}) \delta_{km}
=  (J_{k}-J_{k+1}) \delta_{km},
\label{identities}
\end{equation}
and ${\mathbf K}$ is the $(N-1) \times (N-1)$ Cartan matrix given in  \eqref{Cartanmatrix}.

Solving \eqref{frank1} and inserting it into the grade $0$ projection of 
the zero-curvature equation \eqref{zerocurva} determines the flow of
the model through equation $\partial_t A_0= \partial_{x}  D^{(0)}$, with 
$t \equiv t_2$, that reads 
in vector notation as :
\begin{equation}
\partial_t {\mathbf v}  = \partial_x {\mathbf d}\, . %
\label{vectorrel}
\end{equation}
After multiplying  both sides of relation \eqref{vectorrel}
by the Cartan matrix ${\mathbf K}$ and making use of relation \eqref{frank1} 
we obtain:
\begin{equation}
\begin{split}
\partial_t \sum_{m=1}^{N-1} \left( {\mathbf K}\right)_{km} v_{m}&=
\partial_t \left( 2 v_k - v_{k+1}-v_{k-1}\right)
= \partial_x \left( \partial_x {\mathbf F}  + {\mathbf V} {\mathbf F}
\right)_{k}\\
&=  \partial_x  \bigg\lbrack \partial_x (v_{k+1}-v_{k-1}) + V_{kk}
(v_{k+1}-v_{k-1}) \bigg\rbrack, \;\; k=1, {\ldots} , N-1 \, .
\end{split}
\label{frank2}
\end{equation}
Recalling the identities \eqref{Fkdefs} and  \eqref{identities}
the above equation \eqref{frank2} can be rewritten as:
\begin{equation}
\partial_t   \left( J_{k+1}-J_{k} \right)=
- \partial_x  \bigg\lbrack \partial_x ( J_k+J_{k+1} ) - \left( J_{k+1}-J_{k} \right)
( J_k +J_{k+1}) \bigg\rbrack, \quad k=1, {\ldots} , N-1\, ,
\label{5prime}
\end{equation}
that defines $t_2$ flows of $sl(N)$ mKdV hierarchy in agreement  with the result given earlier in \cite{schiff93}.
From equation \eqref{5prime} one can derive %
a flow for an individual
$J_m$ as
\begin{equation} 
\partial_t J_m =  \partial_x \bigg[ 
  \sum_{k=1}^{N-1} \frac{N-2k}{N} J_{k+m}^\prime 
 +J_m^2- \frac1N \sum_{k=1}^{N} J_k^2 \bigg] 
 , \quad m=1,{\ldots} , N-1 \, .
 \label{patJm}
 \end{equation}

\section{Self-similarity limit and $A^{(1)}_{N-1}$ Painlev\'e Equations}
\label{section:scaling}

Introducing the self-similarity limit through relations
\begin{equation} \label{scaling}
z = t^{-1/2} x ,\qquad J_i (x,t)= t^{-1/2} J_i (z) \, ,
\end{equation}
one obtains in such limit from equations  \eqref{5prime} after an integration     :
\begin{equation}
\left( J_i +J_{i+1} \right)_z = -\frac{z}{2} \left( J_i -J_{i+1} \right)
-   \left( J_i^2 -J_{i+1}^2 \right)+ \gamma_i, \qquad
i=1,{\ldots} ,N\, , 
\label{JiJip1}
\end{equation}
with $\gamma_i$ being integration constants that 
satisfy $\sum_{i=1}^N
\gamma_i=0$ and where we introduced the periodicity condition $J_{N+i}=J_i$.

Define now:
\begin{equation}
j_i(z) = J_i(z) + \frac14 z \, , 
\label{newdefj}
\end{equation}
with the zero trace  condition $\sum_{k=1}^N J_k=0$ replaced by
\begin{equation}
\sum_{i=1}^N j_i =   \frac{N}{4} z \, 
\label{newcondj}
\end{equation}
and impose periodicity condition:
\begin{equation}
j_i =  j_{i+N} \,.
\label{period}
\end{equation}

In this notation and with  $\alpha_i=\gamma_i+\frac12$, such that
$\sum_{i=1}^N \alpha_i= N/2$, equations \eqref{JiJip1} simplify to: 
\begin{equation}
\left( j_i +j_{i+1} \right)_z =  
- \left( j_i -j_{i+1} \right)
\left( j_i +j_{i+1}  \right) + \alpha_i
, \qquad
i=1,{\ldots} ,N \, ,
\label{mape}
\end{equation}
in which we recognize the dressing chain equation  \cite{Veselov-Shabat}.

There is also an alternative way of obtaining the dressing chain
equation \eqref{mape} from the $t_2$-flow  \eqref{5prime} of the
$sl(N)$ mKdV hierarchy that employs the traveling wave reduction:
\begin{equation}\label{xct}
J_i(x,t)=j_i (z) - \frac{C}{2},\qquad z=x-C t\, .
\end{equation}
Here $C$ is a constant representing speed of a wave.
We find that such reduction yields the same differential system
\eqref{mape}, but with the major difference being that the summation 
of $j_i$ now yields a constant:
\begin{equation} \label{nc2}
\sum_{i=1}^N j_i=\frac{NC}{2}
\end{equation}
and therefore the sum of all $N$ equations \eqref{mape} is:
\begin{equation} \label{alpha0}
2\sum^{N}_1 j_{i\,z} =\sum^{N}\alpha_i =0 \, .
\end{equation}
This is the case of $\alpha\equiv \sum^{N}_{i=1} \alpha_i=0$ 
for which the periodic dressing chain
admits a parametric dependent Lax representation \cite{Veselov-Shabat}.
Equation \eqref{mape} with condition \eqref{alpha0} for 
$N=3$ yields $I_{30}$ (denoted as equation XXX  in \cite{ince}) 
as shown in \cite{coalescence}.
See reference  \cite{victorsthesis} for relation between 
equation \eqref{mape} with condition \eqref{alpha0} for 
$N=4$ and $I_{38}$ (equation XXXVIII  in \cite{ince}).

Define  now
\begin{equation}
f_i = j_i+j_{i+1} \, ,
\label{gidef}
\end{equation}
then due to identities shown in Appendix \ref{section:fromj2f} it follows
that equation \eqref{mape}  becomes in this new notation
\begin{equation}
\frac{d f_i}{d z}  =   f_i  
\sum_{r=1}^{k} \big(f_{i+2r-1}  -   f_{i+2r}  \big)   +
\alpha_i , \quad
i=1,{\ldots} ,N \, ,
\label{solut22}
\end{equation}
for  $N=2k+1$ odd  and with conditions $\sum_{i=1}^N f_i =Nz/2$
and $\sum_{i=1}^N \alpha_i =N/2$. Equation \eqref{solut22} reproduces $A^{(1)}_{N-1}$ 
Painlev\'e equations for $N$  odd.
For even $N=2n$ we obtain from \eqref{mape}
\begin{align} 
nz\frac{d f_i}{d z}=&f_i \left(  \sum_{1\leq r\leq s\leq n-1}f_{i+2r-1}f_{i+2s}
-\sum_{1\leq r\leq s\leq n-1}f_{i+2r}f_{i+2s+1}+\sum_{r=1}^{n}\left(1 -2\alpha_{i+2r}\right) 
\right) \nonumber\\
&+nz \alpha_i, \qquad
i=1,{\ldots} ,N \label{evenPeq}
\end{align}
reproducing $A^{(1)}_{2n-1}=A^{(1)}_{N-1}$ 
Painlev\'e equations  with 
conditions %
\begin{equation}
\sum_{r=1}^{n}f_{i+2r}=   \sum_{r=1}^{n}f_{i+2r+1} =\frac{n}{2}z \, .
\label{condsN2na}
\end{equation}

We notice that equations \eqref{mape} are manifestly invariant under order-$N$ automorphisms:
 \begin{equation}
 \pi (j_{i})=j_{i+1} , \qquad  \pi (\alpha_i)= \alpha_{i+1} \, ,
 \label{piJi}
 \end{equation}
 \begin{equation}
 \rho (j_{i})=-j_{N+1-i} , \qquad  \rho (\alpha_i)= -\alpha_{N+1-i} 
 , \qquad  \rho(z)=-z\, .
 \label{rhoJi}
 \end{equation}
In the next section we will  obtain the remaining B\"acklund symmetries by employing the
gauge transformations of the zero-curvature equations.
\section{Gauge transformations of $sl(N)$ mKdV 
zero-curvature equations }
\label{section:gaugetrans}

\subsection{Introducing the gauge transformation, its basic properties
	and parameters}

	Below we will discuss symmetries of equations \eqref{mape}
	derived from gauge transformation of the $\widehat{sl} (N)$ zero-curvature
	equation \eqref{zerocurva}  compactly rewritten as
	\begin{equation} \label{zceq}
	[\partial_x+A_x,\partial_t+A_t]=0\, ,
	\end{equation}
	with
	\[
	A_x= E^{(1)} +A_0\,,\quad \; A_t=   D^{(0)}+ D^{(1)}+D^{(2)}\,.  \]
	The term  $D^{(2)}$ is given in \eqref{D2}, $D^{(1)}$ in \eqref{Done}
	and $D^{(0)}= \sum_{a=1}^{N-1} d_a h_{\alpha_a} $ is such that its matrix elements
	satisfy  the second-flow equation $\partial_t v_i=
	\partial_x d_i$. 
	We will now  discuss the gauge symmetries of
	equation \eqref{zceq} that preserve the matrix form of potentials $A_x, A_t$ and will be shown to reproduce the B\"acklund symmetries of Painlev\'e equations  in the self-similarity limit.
In the setting of \eqref{zceq} we will 
explore invariance of \eqref{5prime} under the gauge transformation
\begin{equation} \label{gaugetransform}
A_\mu(J_i)\to A_\mu (\tilde{J_i}) =U A_\mu (J_i)U^{-1}+
U\partial_\mu U^{-1}, \qquad \mu=x, t
\end{equation}
or 
\begin{equation}\label{ua}
U A_\mu(J_i)=A_\mu(\tilde{J}_i)U+ \partial_\mu U
\end{equation}
where $U=U(\tilde{J},J)$ maps from $J_i$ configuration to 
another one denoted by $\tilde{J}_i$, while preserving the zero-curvature
equation. 

We will work with the gauge transformations that has been previously constructed  to produce B\"acklund transformations in the context of $sl(N)$ mKdV hierarchy \cite{globo} (see also \cite {fordy} for the associated relativistic Toda  model)  to be of the form, 
$U=I + U_{-1}$, where $I$ is the identity matrix and $U_{-1}$ is a
grade $-1$ element of $\widehat{sl}(N)$  :
\begin{equation}
\begin{split}
U(\tilde{J},J) &\equiv
U ( \beta_1, {\ldots} , \beta_N) =\\
&= 
\left(\begin{matrix}
1 & 0 & & \ldots  &\lambda^{-1}\beta_N (x,t) \\
\beta_1(x,t) & 1 & &  \cdots & 0 \\
0  &  \ddots & \ddots &  \cdots     & \\
\vdots &  & \beta_{N-2} (x,t) & 1       &0 \\
0& \cdots & &\beta_{N-1} (x,t)  &1
\end{matrix}\right)
\label{UmatrixN}
\end{split}
\end{equation}
where $\beta_i (x,t), i=1, {\ldots} ,N$  are coefficients parameterizing  grade  $-1$ component of
$\widehat{sl} (N)$ and will be determined  by imposing 
	\eqref{ua} as a graded equation.

For convenience we employed above a notation that expresses explicitly the
dependence of gauge transformation $U$ on the parameters $\beta_i$  
which in turn,
	depend upon the original   and the
	transformed configurations,   $ \{J \}$ and $\{ \tilde{J} \}$ respectively. 

We now consider matrix elements of both sides of relation \eqref{ua}
and derive differential equations for the gauge parameters $\beta_i$ that 
follow from them.
Plugging expression for the matrix $U$ 
into equation \eqref{ua} with $\mu=x$ and considering its diagonal
$(i,i)$ element  we obtain an expression :
\begin{equation}\label{diagjjb}
(i,i):\qquad \tilde{J}_i = J_i -\beta_i+ \beta_{i-1} , \quad i=1, {\ldots}, N
\end{equation}
for the transformed configuration
of $\tilde{J}_i$ in terms of  $J_i$  and the $\beta_i$ coefficients of 
$U ( \beta_1,{\ldots} , \beta_N)$. 
The $(i+1,i)$ element of  \eqref{ua} with $\mu=x$ 
is given by 
\begin{equation}\label{bix}
(i+1,i):\qquad \beta_{i,x}= J_i \beta_i-\tilde{J}_{i+1}\beta_i
=\beta _i \left(-\beta_i+\beta_{i+1}+J_i-J_{i+1}\right), \;i=1,{\ldots}, N-1
\end{equation}
Integrating  \eqref{bix} yields :
\begin{equation}\label{betaxxx}
\beta_1=B_1(t)e^{\int (J_1-\tilde{J_2} )dx},\;\cdots \;,  
\beta_i=B_i(t)e^{\int (J_i-\tilde{J}_{i+1} )dx},
\;\cdots \; , \beta_N=B_N (t) e^{\int J_N-\tilde{J_1} dx}
\end{equation}
where we indicated explicitly that $B_i$-coefficients 
do not depend on $x$ but may depend on the $t$ variable.
However  considering the $(i+1,i)$ element of  \eqref{ua} with $\mu=t$ 
we obtain :
\begin{equation}\label{bit}
(i+1,i):\quad \partial_t \partial_x \ln \beta_{i}= 
\partial_t  (J_i -\tilde{J}_{i+1})\; \to\; 
\beta _i = B_i (x) \exp \int (J_i -\tilde{J}_{i+1}) d x, \;i=1,{\ldots}, N-1
\end{equation}
We thus conclude that all coefficients $B_i$ are constants as is
consequently the  product of all $\beta_i$ :
\begin{equation}\label{bbb}
\beta_1 (x,t) {\cdots} \beta_N (x,t) =B_1  {\cdots} B_N =
\rm{constant}
\end{equation}
which  shows that  there are  in fact only $N-1$ independent   $\beta-$parameters.

For diagonal elements of  \eqref{ua}  for  $\mu=t$ we obtain :
\begin{equation}\label{diagip1}
\begin{split}
 (d_{i+1}-\tilde{d}_{i+1}) -(d_i-\tilde{d}_i ) + \beta_i F_i -\beta_{i+1} \tilde{F}_{i+1} =\,&0,\quad\,
 i=1,{\ldots} ,N-2 \\
 d_1-\tilde{d}_1+\beta_N F_N - \beta_1 \tilde{F}_1 =\,&0 \\
  d_{N-1}-\tilde{d}_{N-1}+\beta_N \tilde{F}_N - \beta_{N-1} F_{N-1} =\,&0
\end{split}
\end{equation}
Next we use that $\partial_x d_i= \partial_t v_i$ and
$\tilde{v}_i=v_i+\beta_N -\beta_i$ with $F_i, F_N$ defined in \eqref{Fkdefs}
and accordingly transforming as $\tilde{F}_i=F_i+\beta_{i-1} -\beta_{i+1}$,
to obtain from \eqref{diagip1} that the quantities 
\begin{equation} \label{gammaidefs}
\Gamma_i \equiv \partial_t \beta_i -\partial_x (\beta_i F_i 
-\beta_{i} \beta_{i+1} ), 
\end{equation} 
are
always equal to each other:
\begin{equation}\label{diagip2}
\Gamma_i= \Gamma_{i+1} , \quad i=1, {\ldots} , N-1 \, .
\end{equation}
A composition law may be derived  by acting successively with $U(\tilde{J},J)$ to transform from 
	$J$ to ${\tilde J}$, followed by ${\tilde J} \, \to \tilde{\tilde J}$ 
	transformation, $U (\tilde{\tilde{J}},\tilde{J})$. It therefore follows 
	the composite 
	gauge transformation
\begin{equation} \label{twous}
U (\tilde{\tilde{J}},J)= U (\tilde{\tilde{J}}, \tilde{J} ) \, U(\tilde{J},J)
\end{equation}

We start by making a remark concerning matrix 
\[ U( \beta_1,{\ldots} ,\beta_{i-1},0,\beta_{i+1},\beta_{i+2}, \beta_{i+3},{\ldots}
,\beta_N )\] with one parameter $\beta_i =0$ being zero. 
As a matrix it can be represented by  an ordered product of
 matrices  $U_i$
		with only  one non-zero  parameter $\beta_i$ :
	\begin{equation}\label{Ubi}
	U_i \equiv  U( 0,{\ldots} ,\beta_i, {\ldots} ,0).
	\end{equation}
	In order to  maintain the form of the gauge matrix as in equation \eqref{UmatrixN}
		the product 
		must be ordered starting  from the right with $U_{i+1}$  ending at 
		$U_{i-1}$ as can be explicitly verified  to be,
	\begin{equation}\label{order}
	U( \beta_1,{\ldots} ,\beta_{i-1},0,\beta_{i+1},\beta_{i+2}, \beta_{i+3},{\ldots}
	,\beta_N )=   U_{i+1} U_{i+2} \cdots U_{N-1}  U_N U_1 U_2 \cdots U_{i-1}.
	\end{equation}
 In the next subsection it will be explained  that the gauge matrices considered in this section must,
		in the self-similarity limit,  have at least one 
		parameter equal to zero.   They  all can be represented
		by an  ordered matrix composition   in terms of a basis of   single gauge 
		matrices $U_j$
		since  the right hand side of equation \eqref{order}
		only involves the matrix multiplication of  single $U_i$ matrices. 
	However each gauge transformation
	not only acts as a matrix but also as a transformation that
	obeys the composition rule \eqref{twous}. To take into consideration
	those two
	actions we will introduce below a ``star''-product  of two gauge 
	matrices \eqref{UiUk}.

 A basis for the B\"acklund transformation  can be proposed  to consist of  generators 
		with only one non-zero coefficient, $\beta_i \ne0, \quad i=1, \cdots N-1$, it 
		holds from the transformation rule \eqref{diagjjb}
		that $U_i$ from \eqref{Ubi} generates:
	\begin{equation}
	U_i %
	:  \;\;
	J_i \to  J_i - \beta_i, \;\; 
	J_{i+1} \to J_{i+1}  +\beta_i, \;\; J_j \to J_j , \, j \ne i, i+1
	\label{JiJip1bi}
	\end{equation}
	with $\beta_i$ that
	satisfies differential
	equations :
	\begin{equation}
	\beta_{i ,\, x}= \beta_i \left(- \beta_i+ J_i-J_{i+1}
	\right)
	\label{1overbid}
	\end{equation}
	and 
	\begin{equation} 
	\partial_t \beta_i = \left( F_{i} \beta_i \right)_x=\left( (J_i+J_{i+1}) \beta_i \right)_x
	\label{t2bi}
	\end{equation}
	due to relations \eqref{bix} and \eqref{diagip2}. Similar relations were derived in \cite{schiff93} by a different method.

\subsection{Gauge transformations in the self-similarity limit
as the extended affine Weyl group symmetries}
\label{subsection:simlimgtrans}
Recall now the self-similarity limit \eqref{scaling} that in accordance with equations \eqref{diagjjb} and \eqref{bix} needs to be  augmented by 
\begin{equation} \label{scalingb}
\beta_i (x,t)= t^{-1/2} b_i (z)\ .
\end{equation}
The self-similarity limit can not be applied consistently to the  constant in \eqref{bbb}
as long it is different from zero since the self-similarity limits of the right 
and left hand side of equations 
\[
\partial_x (\beta_1 (x,t) {\cdots} \beta_N (x,t) )=
\partial_t (\beta_1 (x,t) {\cdots} \beta_N (x,t) )=0
\]
will lead to inconsistent limits  if all $\beta_i (x,t) \ne 0$.
Thus at least one of the $B_i$ constants will need to vanish for us to be able 
to take the self-similarity limit.
Clearly in such case $\Gamma_i=0$ in \eqref{diagip2} and this equation
yields in the scaling  limit 
the following algebraic relations : 
\begin{equation}\label{diagip2s}
b_i b_{i+1} -b_i (j_i+j_{i+1} ) +\kappa_i=b_i b_{i+1} - b_i f_i +\kappa_i=0
, \quad i=1, {\ldots} , N-1
\end{equation}
where $\kappa_i$ are integration constants.
Furthermore equation \eqref{diagjjb} becomes in the self-similarity limit
\begin{equation}\label{diagjjbs}
\tilde{j}_i = j_i -b_i+ b_{i-1} \quad \to \quad 
\tilde{f}_i = f_i
-b_{i+1}+ b_{i-1} 
, \quad i=1, {\ldots}, N
\end{equation}
and equation \eqref{betaxxx} becomes
\begin{equation}\label{betaxxxs}
b_1=b^{(0)}_1 e^{\int j_1-\tilde{j_2} dz},\;\cdots \;,  
b_i=b^{(0)}_i e^{\int j_i-\tilde{j}_{i+1} dz},
\;\cdots \; , b_N=b^{(0)}_N  e^{\int j_N-\tilde{j_1} dz}
\end{equation}
In the self-similarity limit the choice \eqref{Ubi} results in 
constants $b^{(0)}_i\ne 0$ and $b^{(0)}_j =0, j \ne i$, and we 
obtain  from relations \eqref{diagip2s} that 
\begin{equation} \label{diagip2ss} 
\kappa_i= b_i f_i \quad \to \quad b_i=\frac{\kappa_i}{f_i}
\end{equation} 
Let us recall equation \eqref{t2bi} describing the time flow of $\beta_i$
parameter.
According to the self-similarity limits  \eqref{scaling}  and  \eqref{scalingb}  
we obtain from \eqref{t2bi} :
\begin{equation} \label{t2bis} \left( (J_i+J_{i+1} + z/2)b_i \right)_z=0 \;\; \to \; \;
1/b_i (z) = (J_i(z) +J_{i+1} + \frac12 z)/\kappa_i= (j_i+j_{i+1}
(z))/\kappa_i \, ,
\end{equation}
where we have chosen the integration constant $\kappa_i$ to be in agreement
with \eqref{diagip2ss}.
We now turn our attention to equation \eqref{1overbid} that becomes in
the self-similarity limit 
\begin{equation} \label{1overbids}
b_{i\, z}= b_i ( - b_i +j_i- j_{i+1}), \quad \to \quad \partial_z \ln
b_i +b_i = j_i- j_{i+1}
\end{equation}
Plugging relation $ b_i = \kappa_i /f_i$ from equation \eqref{t2bis}
into equation \eqref{1overbids} one obtains 
\[
\partial_z \ln \left( \frac{\kappa_i}{f_i} \right) +
\frac{\kappa_i}{f_i} = j_i- j_{i+1}
\]
that agrees with the dressing equation \eqref{mape} for
$\kappa_i=\alpha_i$.
Indeed it is easy to explicitly  verify that the dressing chain 
equation is invariant under the symmetry 
operations 
\begin{equation}
\begin{split}
 j_i &\stackrel{U_i}{\longrightarrow} \tilde{j}_i= j_i-b_i=  j_i - \frac{\kappa_i}{j_i
+j_{i+1}},\quad
j_{i+1}  \stackrel{U_i}{\longrightarrow} \tilde{j}_{i+1}= j_{i+1} +b_i= j_{i+1} + \frac{\kappa_i}{j_i
+j_{i+1}}, \\
j_k &\stackrel{U_i}{\longrightarrow} j_k, \; k \ne i,
k\ne i+1
\end{split} \label{Tinewj}
\end{equation}
generated according to transformation rule \eqref{JiJip1bi}  by
self-similarity limit counterpart of quantity defined in \eqref{Ubi}
for $\kappa_i=\alpha_i$, when it is accompanied by transformations
of coefficients $ \alpha_i \to
-\alpha_i,  \alpha_{i \pm 1}\to \alpha_{i \pm 1} +\alpha_i$.
We will derive below these transformations directly from actions of
$U_i$ matrices (see equations \eqref{Ubialphai} and \eqref{sip1def}).
Thus the above transformation induced by the gauge matrix $U_i $
from relation \eqref{Ubi}
agrees with known B\"acklund
transformations of $A^{(1)}_{N-1}$ Painlev\'e equations
\cite{adler93}.

We will need to define the composition laws for the gauge transformations 
that will hold when acting with an additional gauge transformation $U (\tilde{\tilde{J}},\tilde{J})$ 
following $U(\tilde{J},J)$ as shown in \eqref{twous}. 
The ordering of equation \eqref{twous} will be important when
defining below products of $U_i$ from equation \eqref{Ubi}. 
We now define a ``star''-product  of two single $U_i$ and 
$U_k$ matrices that follows the composition rule \eqref{twous} :
\begin{equation}\label{UiUk}
 U_i (b_i) \star  %
U_k(b_k )= U_i (\tilde{b}_i) \, U_k (b_k) \, .
\end{equation}
Note that in addition to matrix multiplication between $U_i$ and 
$U_k$ (on the right hand side
of the above equation), %
$U_k$  simultaneously
acts  through transformation of all 
the quantities $b_i,j_i$ to 
$\tilde{b}_i,\tilde{j}_i$ according to 
\begin{equation} 
 j_i \stackrel{U_k}{\longrightarrow}{\tilde j}_i=  U_k (j_i)=
\begin{cases}
j_i \to j_i  & i\ne k, i \ne k+1\\
j_i \to j_i-b_i & i=k\\
j_{i} \to j_{i} +b_{i-1}& i=k+1
\end{cases}
\label{UkonUI}
\end{equation}
where we used transformation rule \eqref{diagjjbs}.
The following commutation relation follows from \eqref{UkonUI}
and from the corresponding rules of matrix multiplication for matrices $U_i$
and $U_k$ that do not have any neighboring matrix element $b_i,b_k$: 
\begin{equation} \label{UiUkcommute}
U_i (b_i) \star  %
U_k(b_k ) = U_k(b_k ) \star U_i (b_i)
, \qquad  i\ne k, i \ne k \pm 1
\end{equation}

We now consider the ``star''-product \eqref{UiUk} with
$k=i$ meaning we let $U_i( b_i)$ act twice with each $U_i$ acting simultaneously
through matrix
multiplication and through transformation of all the quantities $b_k,j_k$ to 
$\tilde{b}_k,\tilde{j}_k$. Thus the result of performing the gauge
transformation twice is $U_i (b_i)
\star   U_i(b_i) $ or explicitly :
\[
 U_i (b_i) \star  %
U_i(b_i)= U_i (\tilde{b}_i) \, U_i (b_i)
\,  ,
\]
where $\tilde{b}_i=U_i (b_i)$ must satisfy equation
\begin{equation} \label{1overbids2}
\tilde{b}_{i\, z}= \tilde{b}_i ( - \tilde{b}_i +(j_i-b_i)- (j_{i+1}+b_i))=
\tilde{b}_i ( - \tilde{b}_i +j_i- j_{i+1}-2b_i)
\end{equation}
obtained from equation \eqref{1overbids} by acting with $U_i$.
Plugging in the above expression equation \eqref{1overbids} that  amounts to setting  
$j_i- j_{i+1}=\partial_z \ln b_i +b_i $ we get:
\begin{equation} \label{tbb}
\partial_z \ln \tilde{b}_i +\tilde{b}_i= \partial_z \ln b_i -b_i\;\to \;
\partial_z \ln \frac{\tilde{b}_i}{b_i} =-\tilde{b}_i -b_i
\end{equation}
Since $\frac{\tilde{b}_i}{b_i} = \frac{\tilde{\kappa}_i}{\kappa_i} $
because $\tilde{f}_i=f_i$, 
the left hand side of the last equation in \eqref{tbb} is zero and
consequently
\[
\tilde{b}_i = - b_i
\]
and $\tilde{\kappa}_i=-\kappa_i$ or $\tilde{\alpha}_i=-\alpha_i$. It follows now by  simple multiplication
\begin{equation} \label{Uisquare}
 U_i \star U_i= U( 0,{\ldots},0 , -b_i, 0,{\ldots} 0)\,
 U( 0,{\ldots},0 , b_i, 0,{\ldots} 0)= 1
\end{equation} 
with the identity matrix on the right hand side of the above equation. We thus obtain that $U_i \star U_i=1$ %
for all $i=1,2,{\ldots} ,N$. We also learn that
\begin{equation} \label{Ubialphai}
\alpha_i \stackrel{U_i}{\longrightarrow} -\alpha_i
\end{equation}
Next we will verify that 
\begin{equation}\label{Uiip1}
U_{i\, i+1} \equiv U( 0,{\ldots},0 , b_i,b_{i+1}, 0,{\ldots} 0)=
U( b_i) \star U( b_{i+1}) 
\end{equation}
according to the definition given in relations \eqref{UiUk} and \eqref{UkonUI}.
Therefore we  now  consider the $U$-matrix from \eqref{UmatrixN}
with two neighboring $\beta_i, \beta_{i+1}$ that
are different from zero and all the other $\beta_j=0$. The following
equations hold for $\beta$'s and their self-similarity limits :
\begin{equation} 
\begin{split}\label{neighboring}
\beta_{i ,\, x}&= \beta_i \left(- \beta_i+\beta_{i+1} + J_i-J_{i+1}
\right)\quad \to \quad \partial_z \ln
b_i +b_i = b_{i+1} +j_i- j_{i+1},\\
\beta_{i+1 ,\, x}&= \beta_{i+1} \left(- \beta_{i+1} + J_{i+1}-J_{i+2}
\right) \quad \to \quad \partial_z \ln
b_{i+1} +b_{i+1} = j_{i+1}-j_{i+2}
,\\
 \partial_t \beta_i &= \left( \beta_i F_{i} -\beta_i\beta_{i+1}
 \right)_x   \quad \to \quad b_{i}= \frac{\kappa_{i}}{f_i-
 \kappa_{i+1}/f_{i+1}},\\
  \partial_t \beta_{i+1} &= \left( \beta_{i+1} F_{i+1} \right)_x
  \quad \to \quad b_{i+1}= \frac{\kappa_{i+1}}{f_{i+1}},
\end{split} 
\end{equation} 
which agree with the dressing equations \eqref{mape} for $f_i, f_{i+1}$ for
$\kappa_{i+1}=\alpha_{i+1}, \kappa_{i}=\alpha_i +\alpha_{i+1}$.
Comparing expression for $b_i$ given in \eqref{neighboring} with those
in \eqref{UkonUI} that define the $\star$-product \eqref{UiUk} we
find that the relation \eqref{Uiip1} holds.

Further it follows that $U_{i\, i+1}$, as defined in \eqref{Uiip1},  
with  the non-zero neighboring $U$-matrix elements  $b_i,
b_{i+1}$  will induce the following non-zero transformations
\begin{equation} \label{Ubiip1trans}
\begin{split}
f_i &\to \tilde{f}_i=f_i+\cancel{b_{i-1}}-b_{i+1}=
 f_i-\frac{\alpha_{i+1}}{f_{i+1}}, \\
  f_{i+1} &\to
 \tilde{f}_{i+1}=f_{i+1} + b_i- \cancel{b_{i+2}} =f_{i+1} +
 \frac{\alpha_i+\alpha_{i+1} }{f_i-\frac{\alpha_{i+1} }{f_{i+1} }}, \;\;
 \\
f_{i+2} &\to \tilde{f}_{i+2}= f_{i+2} +b_{i+1}-\cancel{b_{i+3}}=f_{i+2} +
 \frac{\alpha_{i+1} }{f_{i+1} }
\, , \\
f_{i-1} &\to
 \tilde{f}_{i-1}=f_{i-1} + \cancel{b_{i-2}}- b_{i} 
=f_{i-1} -\frac{\alpha_i+\alpha_{i+1} }{f_i-\frac{\alpha_{i+1}
}{f_{i+1}}},
 \end{split} \end{equation}
where we crossed out those coefficients $b_i$'s that are absent  in the $U_{i\, i+1}$
matrix.

If we now define the  $s_{i+1}$ transformation as :
\begin{equation} \label{sip1def}
s_{i+1}: f_i\to {\tilde f}_i= f_i - \frac{\alpha_{i+1}}{f_{i+1}}, \,f_{i+1} \to {\tilde f}_{i+1}=f_{i+1}, \,
f_{i+2} \to {\tilde f}_{i+2}=f_{i+2}+\frac{\alpha_{i+1}}{f_{i+1}},\,
\end{equation}
together with $ s_{i+1}: \alpha_i \to {\tilde
	\alpha}_i = \alpha_i+\alpha_{i+1}$, 
then we can represent the transformation \eqref{Ubiip1trans} 
as a result of action of $s_{i+1}$ 
followed by $s_i$ transformation:
\[
\begin{split}
s_i: {\tilde f}_i&\to {\tilde f}_i= f_i - \frac{\alpha_{i+1}}{f_{i+1}}, 
\,{\tilde f}_{i+1} \to {\tilde f}_{i+1}+\frac{{\tilde
\alpha}_i}{{\tilde f}_i} = f_{i+1}+  \frac{\alpha_i+\alpha_{i+1}}{f_i -
\frac{\alpha_{i+1}}{f_{i+1}}}\\
{\tilde f}_{i+2} &\to {\tilde f}_{i+2}
=f_{i+2}+\frac{\alpha_{i+1}}{f_{i+1}} \; , 
{\tilde f}_{i-1} \to {\tilde f}_{i-1} -
\frac{\tilde{\alpha}_i}{\tilde{f}_i}=
f_{i-1}-\frac{\alpha_i+\alpha_{i+1}}{f_i -
\frac{\alpha_{i+1}}{f_{i+1}}}
\end{split}
\]
Note also that as follows from relation \eqref{Ubialphai} $s_i
(\alpha_i)=-\alpha_i$.
Thus the gauge transformation $U_{i\, i+1}  $ 
agrees with the composite $s_i s_{i+1}$ B\"acklund transformation and
it follows from this identification by repetitive action with $s_is_{i+1}$ that
\begin{equation}
\label{Uii1cube}
(U_i \star U_{i+1} )^3=1 \, .
\end{equation}

More generally the gauge transformation 
$U(0,{\ldots}
,0,b_i,b_{i+1},{\ldots} ,b_{i+k},0,{\ldots},0 )$
can be factorized into a star-product of one parameter gauge transformations $U_i$  that 
corresponds to the composite $s_is_{i+1}{\ldots} s_{i+k}$ 
B\"acklund transformation with
relations between $b_i$'s being of the form (for $k=3$) :
\[
b_i=\frac{\kappa _i}{\frac{\kappa _{i+1}}{\frac{\kappa_{i+2}}{\frac{\kappa _{i+3}}
{b_i-j_i-j_{i+3}}-j_{i+2}-j_{i+3}}-j_{i+1}-j_{i+2}}-j_i-j_{i+1}}
\]
 and so on for higher $k$.

The gauge matrices $U(0,{\ldots} ,b_i,b_{i+1},{\ldots}
,b_{i+n},{\ldots},0 )$ and $U(0,{\ldots} ,b_j,b_{j+1},{\ldots}
,b_{j+m},{\ldots},0 )$ such that $j+m< i-1$ or $i+n< j-1$ will commute
leading to commutation of the B\"acklund transformations
$s_i \cdots s_{i+n}$ with $s_j \cdots s_{j+m}$ under the same conditions.

The key conclusion of this section is that the identities \eqref{UiUkcommute}, \eqref{Uii1cube} and \eqref{Uisquare} 
establish equivalence of the group of gauge transformation endowed 
with the $\star$-product to the extended affine Weyl group $A^{(1)}_{N-1}$
with its  fundamental relations
\[ s_i^2=1, \; (s_is_{i+1})^3 =1,  \; s_i s_k=s_k s_i ,\; k\ne i \pm 1
\]
that agree with \eqref{UiUkcommute}, \eqref{Uii1cube} and \eqref{Uisquare} .

For the remaining part of this subsection we will discuss how that gauge 
transformations will function
in case of the traveling wave limit \eqref{xct}. In this limit  we obtain 
\begin{gather}
\partial_x(\beta_1 \cdots \beta_N)\to \partial_z(b_1 \cdots b_N)\\
\partial_t(\beta_1 \cdots \beta_N)\to -C\partial_z(b_1 \cdots b_N)
\end{gather}
and we conclude that 
\begin{equation} \label{zb1nz}
\partial_z (b_1 \cdots b_N) =0
\end{equation} 
Contrary to the case of the self-similarity limit, the condition
$b_1 \cdots b_N=0$ is no longer the only remaining option. 
From equations \eqref{1overbid} we find 
\begin{equation}
 b_{i ,\, z}= b_i \left(- b_i+ j_i-j_{i+1}
\right)
\label{1overbidtravel}
\end{equation}
and combined with relation \eqref{diagip2s} we find that the
condition \eqref{zb1nz} has another valid solution apart from %
$b_1 \cdots b_N=0$, namely that $\kappa_1 \cdots \kappa_N=0$ while $b_i\ne 0, i=1,...,N$.
Let us set $\kappa_{i-1}=0$, then   from relations \eqref{diagip2s}
we find a solution: 
\[
b_{i-1}=f_{i-1},\; b_{i+1}=f_i-\frac{\kappa_i}{f_{i-1}},\;
b_{i+2}=f_{i+1}-\frac{\kappa_{i+1}}{f_i-\frac{\kappa_i}{f_{i-1}}}
\]
etc. Thus in this case we can obtain a solution that has all 
the coefficients $b_i\ne 0$. Will such transformation still be
equivalent to a finite product of B\"acklund transformations $s_i$?
To answer this question let us consider for illustration the case of $N=3$ and set $\kappa_1=0$. Then the
relevant $3 \times 3$ matrix $U (b_1,b_2,b_3)$ has arguments:
\begin{equation}
b_2=f_1,\qquad b_3= \frac{\kappa_2}{f_1}+f_2,\qquad b_1=
\frac{\kappa_3}{\frac{\kappa_2}{f_1}+f_2}+f_3
\end{equation}
Recalling that the $b_i$ transform the $j_i$ according to relation
\eqref{diagjjbs}   
we find
\begin{equation} \label{N3traveltransf}
\tilde{f}_1= \pi (f_1 +\frac{\alpha_3}{f_3}), \;
\tilde{f}_2= \pi \left(f_2 -\frac{\alpha_3}{f_3}+\frac{\alpha _1+\alpha
_2}{\frac{\alpha_3}{f_3}+f_1} \right),\; 
\tilde{f}_3 = \pi \left(f_3-+\frac{\alpha _1+\alpha
_2}{\frac{\alpha_3}{f_3}+f_1} \right),\; 
\end{equation}
using that
\begin{equation}
\kappa_2=\alpha_1,\quad \kappa_3=\alpha _1+\alpha _2
\end{equation}
as obtained from plugging  the above relations directly back into 
\eqref{mape}.
We also get the transformation on the constants $\alpha_i$ of
\eqref{mape}
\begin{equation}\label{newalpha}
\alpha_1\to-\alpha _1-\alpha _2,\qquad \alpha_2\to 2 \alpha _1+\alpha _2+\alpha _3,\qquad \alpha_3\to \alpha _2
\end{equation}
We see that despite of having all the three $b_i$-coefficients different from
zero that the transformation $U (b_1,b_2,b_3)$ 
obtained in the traveling wave reduction 
is equivalent to 
the composition of B\"acklund transformations $\pi s_3s_1$  that this time involves the automorphism $\pi$ in addition to transformations $s_i$. Thus this example  confirms  that the conclusions
obtained in our study of gauge transformations in the self-similarity
limit extend mutatis mutandis to the traveling wave reduction.

\section{B\"acklund transformations for higher flows}
\label{section:highertimes}

In references \cite{jfg2015,jfg2016} it was observed 
that the 
B\"acklund symmetries obtained out of the gauge symmetries of the zero-curvature equations will hold for all flows  of the integrable model. We can here extend this observation to  all higher-order Painlev\'e equations that are derived  from these  flows. The point we are making here is that the affine extended $A^{(1)}_{N-1}$ Weyl group is the symmetry group for all Painlev\'e equations that are derived by appropriate similarity transformations  from the higher flows of soliton equations. For simplicity we will illustrate this point 
for $N=2 $  mKdV hierarchy with the  commuting flows $t_{2n+1}, n=1,2,3, \ldots$. Here we focus  at the first three higher times $t_3, t_5, t_7$. For all the flows  the $A_x$ matrix remains  defined by:
\begin{equation}
A_x=\left(
\begin{array}{cc}
v  & \lambda \\
1 & - v \\
\end{array} 
\right)\, .
\end{equation}
The gauge matrix $U$ that interpolates between two solutions of the same  mKdV  model  is given by
\begin{equation} \label{UmatrixN2}
U=\left(
\begin{array}{cc}
1  & \frac{\beta_2}{\lambda} \\
\beta_1 & 1 \\
\end{array} 
\right) \, .
\end{equation}
 In the self-similarity limit $U$ generates B\"acklund transformations as described in the previous section.

Following the steps similar to those taken  in Section \ref{section:curvature} we obtain in the zero-curvature framework    the corresponding mKdV equations for the flows  $ t_3, t_5, t_7$ :
\begin{align}
4 v_{t_3}&= v_{3 x}-6 v^2 v_x \, ,\label{mkdv} \\
16v_{t_5}&=30 v^4 v_x-10 v^2 v_{3 x}-40 v v_x v_{2 x}-10 v_x^3+v_{5 x} \, ,
\label{m5}\\
64v_{t_7}=&-140 v^6 v_x+70 v^4 v_{3 x}+560 v^3 v_x v_{2 x}+420 v^2 v_x^3-14 v^2 v_{5 x}\label{m7}\\
&-140 v v_{2 x} v_{3 x}-84 v v_x v_{4 x}-182 v_x v_{2 x}^2-126 v_x^2 v_{3 x}+v_{7 x}
\, .
\nonumber
\end{align}
The usual dimensional considerations \cite{flaschka} lead to the following  self-similarity limits for $t_j=t_3, t_5, t_7$: 
\begin{equation}\label{tjsim}
v(x,t)=\frac{v(z)}{t_j^{1/j}},\qquad \beta_i=\frac{b_i(z)}{t_j^{1/j}},\qquad z=\frac{x}{t_j^{1/j}},\quad j=3,5,7 \,. 
\end{equation}
Taking  such limit of equation \eqref{mkdv} (and integrating once) we get the $P_{II}$ equation:
\begin{equation}
v_{2z}(z)=\alpha +2 v^3 (z)-\frac{4}{3} z v(z)\, ,
\end{equation}
where $v_{i z }$ denotes the $i$-th derivative with respect to the argument $z$.
From the gauge transformation $U$ we inherit  two  B\"acklund transformations. The first  is given by 
 
\begin{equation}
v\to v +b,\qquad b=\frac{2-3 \alpha }{-3 v_z (z)-3 v^2 (z)+2 z},\qquad \alpha\to\frac{4}{3}-\alpha \, ,
\end{equation}
and squares to identity  as follows by inspection.

 Since %
\begin{equation} \label{vminusv}
v\to-v,\qquad\alpha\to-\alpha \, ,
\end{equation}
is an obvious  %
symmetry of $P_{II}$ (and mKdV)  one easily   sees that $\bar{b} (v,\alpha)=b(-v,-\alpha)$ will  be the second  B\"acklund transformation of $P_{II}$ as can also be derived directly from the gauge symmetry argument involving the matrix $U$ \eqref{UmatrixN2}.
Such  symmetry \eqref{vminusv} is also present for  all higher times  $t_n$, therefore it will suffice for us to show from now on only one of the two B\"acklund transformations.

For the $t=t_5$ mKdV equation \eqref{m5} 
after imposing  the appropriate similarity limit from \eqref{tjsim} and performing an integration, we obtain:
\begin{equation}
v_{4 z}=10v^2v_{2 z}+10v v_z^2-6v^5-\frac{16}{5}z v+\sigma
\end{equation}
This equation can be identified with the F-XVII equation of reference  \cite{cosgrove} for the parameters
\[\delta=0,\qquad \lambda=-\frac{16}{5},\qquad \mu=0 \, .\]

As before we are able to derive two Bäcklund transformations from the gauge symmetry argument that are related by the symmetry operation \eqref{vminusv}. We show one of them here:
\begin{equation}
v\to v +b,\qquad b=\frac{8-5 \alpha }{15 v^4-5 v_{3 z}-10 v v_{2 z}+5 v_z^2+30 v^2 v_z+8 z},\qquad \alpha\to\frac{16}{5}-\alpha \, .
\end{equation}
Finally for the $t=t_7$ and the corresponding mKdV flow \eqref{m7} 
the appropriate self-similarity limit from \eqref{tjsim} yields 
\begin{equation}
\begin{split}
v_{ 6 z }&=\alpha +20 v^7+42 v v_{ 2 z}^2-70 v^4 v_{2 z} +14 v_{4 z } v^2-140 v^3 v_z^2+56 v_{3 z }  v_z v\\
&+70 v_z^2 v_{2 z} -\frac{64 z v}{7} \, ,
\end{split}
\end{equation}
after  performing an integration.
One of the 
 two B\"acklund transformations we find in this case is :
\begin{equation}
v\to v +b,\qquad b=\frac{32-7 \alpha }{\text{denom}},\qquad \alpha\to\frac{64}{7}-\alpha \, ,
\end{equation}
where
\begin{align*}
\text{denom}=&-70 v^6-7 v_{5 z }-7 v_{2 z}^2+70 v_z^3-210 v^4 v_z+140 v^3 v_{2 z }+\\
&+14 v_{3 z } v_z +14 v \left(20 v_z  v_{2 z}-v_{4 z }\right)+70 v^2 \left(v_{3 z }+v_z^2\right)+32 z\, .
\end{align*}
Common for all these higher-order Painlevé equations derived from higher  mKdV equations is invariance under the extended affine Weyl group obtained from the B\"acklund gauge symmetries in the self-similarity limit.

\section{Outlook}
\label{section:outlook}
In the recent publications \cite{coalescence,p3-p5s,p3-p5}
we advanced a notion of symmetry being a relevant measure of integrability by 
proposing models that  only kept a part of the affine extended Weyl group symmetry while
maintaining (not violating) the Painlev\'e property. 
Along the same line of the investigation as presented in \cite{coalescence}  we can here propose a deformation of the dressing chain \eqref{mape} 
given for $N=3$ by
\begin{equation}\begin{split}
		\left( j_1 +j_{2} \right)_z &=  
		- \left( j_1 -j_{2} \right)
		\left( j_1 +j_{2}  \right) + \alpha_1 + \eta \left(j_1+j_3
		\right) \\
		\left( j_2 +j_{3} \right)_z &=  
		- \left( j_2 -j_{3} \right)
		\left( j_2 +j_{3}  \right) + \alpha_2 - \eta \left(j_1+j_3
		\right) \\
		\left( j_1 +j_{3} \right)_z &=  
		- \left( j_3 -j_{1} \right)
		\left( j_1 +j_{3}  \right) + \alpha_3 
		\label{deformdressingj2}
	\end{split}
\end{equation}
This model is still manifestly invariant under the transformation
generated by $U_3$
from equation \eqref{Tinewj} such that
\begin{equation}%
	j_3 \to j_3+\frac{\kappa}{j_1+j_3},\quad j_1\to
	j_1-\frac{\kappa}{j_1+j_3}, \;\;j_2 \to j_2 \, ,
	\label{leftoversym}
\end{equation}
with $\kappa=- \alpha_3$ and $\alpha_i \to \alpha_i +  \alpha_3, i=1,2$
although no longer invariant under $U_i, i=1,2$ from \eqref{Tinewj}. Also, the automorphism
$\pi$ does not keep the equations \eqref{deformdressingj2} invariant anymore.
Instead equations \eqref{deformdressingj2}
are invariant under a new $\pi_2$ automorphism \cite{coalescence}
\[
\pi_2 : j_1 \to -j_3, \; j_3 \to -j_1, \; j_2 \to - j_2, \; 
\alpha_1 \to -\alpha_3, \; \alpha_3 \to -\alpha_1, \; \alpha_2 \to - \alpha_2,
\;\eta \to -\eta\, ,
\]
with the transformed $j_i, \alpha_i, i=1,2,3$ now satisfying the
conditions $\sum_{i=1}^3 j_i= -3 z /4$, $\sum_{i=1}^3 \alpha_i= -3/2$ that differ
from the condition \eqref{newcondj} by the minus sign.

We can also run Kovalevskaya-Painlev\'e test as in 
references \cite{Veselov-Shabat,coalescence}
by first assuming that solutions of 
\eqref{deformdressingj2}  equations have to be of the form 
\begin{equation}
	j_i = \frac{a_i}{z} + b_i + c_i z + d_i z^2 + e_i z^3 + \cdots, \qquad
	i=1,2,3 \, ,
	\label{fipoles}
\end{equation}
with the pole chosen at $0$.
After substituting the above equation back into equations \eqref{deformdressingj2} 
we find that coefficients of the first term must satisfy relations $ a_i
+a_{i+1} -a_i^2+a_{i+1}^2=0, i=1,2,3$. For the solution $a_1=1,a_2=0,
a_3=-1$ and any value of $\eta$ we find that $b_2$ and $d_2$ (or $d_1$) will remain undetermined
while all the other coefficients will be fixed. This is a general feature which 
occurs independently of whether the $\eta$ terms are present. 
Thus the pole solutions of the type shown in \eqref{fipoles} will
depend on three parameters ($b_2, d_2$ and position of the pole). This
agrees with integrability property derived from the Kovalevskaya-Painlev\'e test
of the $N=3$ dressing chain and shows that the $\eta$-deformation
preserving the symmetry \eqref{leftoversym} also results in passing
the integrability test.

The issue of finding an underlying integrable model behind the deformed dressing chain 
\eqref{deformdressingj2} remains under investigation.

\appendix

\section{Basic facts about $sl(N)$ algebra }
\label{section:roots}
Since $sl(N)$ Lie algebra  is isomorphic to 
the algebra of trace-less $N \times N$ matrices, we find it convenient  
to work with a matrix representation of
$sl(N)$ with basic matrices being unit upper diagonal matrices 
$E_{\alpha_a}, E_{\alpha_a+\alpha_{a+1}}, {\ldots} , E_{\alpha_a+{\ldots} +\alpha_b}, \; a,b = 1,{\ldots} , N-1$ defined
by their non-vanishing unit matrix elements :
\begin{equation}
\left( E_{\alpha_a}\right)_{ij}=\delta_{i,a}\delta_{j, a+1}, \;\; 
\left( E_{\alpha_a+\alpha_{a+1}}\right)_{ij}=\delta_{i,a}\delta_{j, a+2},  \;
{\ldots} , \;
\left( E_{\alpha_a+{\ldots} +\alpha_b}\right)_{ij}=\delta_{i,a}
\delta_{j, b+1}
\label{Eabij}
\end{equation}
and their corresponding lower diagonal conjugated matrices $E_{-\alpha_a} =
E^\dagger_{\alpha_a}$ etc, with $i,j=1, {\ldots} ,N$.
The so-called Cartan subalgebra of ${\cal G}$ is generated by 
trace-less combinations of the $N \times N$ 
diagonal matrices $h_{\alpha_a}= h_{e_a} - h_{e_{a+1}}$, $a=1, {\ldots} , N-1$ 
with :
\[(h_{e_a})_{ij}  = \delta_{i,a} \delta_{j,a} \, .
\]

It is often convenient to relate the eigenvalues of Cartan subalgebra 
matrices $h_{\alpha_a} $ obtained when acting on $E_{\alpha_a}$ matrices 
to $N-1$ simple roots of $sl(N)$. 
This is facilitated by introducing 
$N$ vectors 
$e_a$ %
obtained as follows.
Let $\{ \hat{e}_1 , {\ldots} ,  \hat{e}_N \}$ be an orthonormal basis in
$\mathbb{R}^N$. Define  $N$ vectors
\[
e_i=  \hat{e}_i - \frac{1}{N} \sum_{j=1}^N  \hat{e}_j, 
\]
that by definition satisfy 
\[ \sum_{j=1}^N e_j=0
\]
and have inner product 
\begin{equation} \label{inner}
\langle e_i , e_j \rangle = \delta_{ij} - \frac{1}{N} \, .
\end{equation}
Define now $N-1$ simple roots of $sl(N)$ as 
\begin{equation} \label{simpleroots}
\alpha_{i}= e_i -e_{i+1}, \quad i=1, {\ldots} ,N-1
\end{equation}
Due to the inner product \eqref{inner}
we obtain the $(N-1) \times (N-1)$ Cartan matrix $K_{ij}$ :
\begin{equation} \label{Cartanmatrix}
K_{ij} = \langle \alpha_i , \alpha_j \rangle = \left\{ \begin{matrix}
2 & i=j \\
-1 & i=j \pm 1\\
0 & \rm{otherwise}
\end{matrix}
\right.
\end{equation}
for products of simple roots defined in \eqref{simpleroots}.

For fundamental weights defined as 
\begin{equation} \label{fundweights}
\Lambda^i =\sum_{j=1}^i e_j,  \quad i=1, {\ldots} ,N-1
\end{equation}
it holds that 
\[
\langle \Lambda^i , \alpha_j \rangle = \delta_{ij} \, .
\]
The Cartan matrix transforms the basis given in term of
fundamental weights to the basis given in terms of roots according to:
\[
\sum_{j=1}^{N-1} \, K_{ij} \Lambda^j= e_i-e_{i+1} = \alpha_i
\]
The Weyl group of sl(N) is a permutation of $N$ vectors $e_i$.

We are interested in generalizing the above structure 
to the loop algebra $ \widehat{sl}(N)$
spanned by $\lambda^n E_{\pm(\alpha_a+{\ldots}
+\alpha_b)}$, $\lambda^n h_{\alpha_a}$ on which we define  the principal gradation
operator:
\begin{equation}
Q = N \lambda \frac{d}{d \lambda} + %
H_{\Lambda}
\label{gradeoperator}
\end{equation}
where 
\[ \Lambda= \sum_{i=1}^{N-1} \Lambda^i\]
and $H_{\Lambda}$ acts by an adjoin action.
One proves that:
\[
\Lambda = 
(N-1) e_1 + (N-2) e_2 +{\ldots} +
e_{N-1}
\]
This gives for $N=3,4,5$
\[
\begin{split}
N=3 : \quad  &\Lambda^1+ \Lambda^2= 2 e_1+e_2= \alpha_1+\alpha_2\\
N=4 : \quad  &\Lambda^1+ \Lambda^2+\Lambda^3 = 3 e_1+2e_2+e_3=
3\alpha_1+4\alpha_2+3 \alpha_3\\
N=5 : \quad  &\Lambda^1+ \Lambda^2+\Lambda^3 +\Lambda^4= 4 e_1+3 e_2+2
e_3+e_4\\
&=2  \alpha_1+3 \alpha_2
+3 \alpha_3+2 \alpha_4
\end{split}
\]
The above  expression  allows an alternative simple expression for $Q$ as:
\begin{equation}
Q = N \lambda \frac{d}{d \lambda} + (N-1) h_{e_1} + (N-2) h_{e_2} +{\ldots} +
 (N-i) h_{e_i}  +{\ldots} + h_{e_{N-1}}
\,,
\label{gradoperator}
\end{equation}
with 
\[
Q_{N=3}= 3 \lambda \frac{d}{d \lambda} +h_{\alpha_1}+h_{\alpha_2},\;\;
Q_{N=4}= 4 \lambda \frac{d}{d \lambda} +3 h_{\alpha_1}+4 h_{\alpha_2}
+3 h_{\alpha_3}\, ,
\]
etc.
One fundamental feature that is of great relevance for the formalism
we present is the fact that loop algebra 
$\widehat{sl}(N)$ decomposes into different grade sectors according to 
their eigenvalues under $Q$:
\begin{eqnarray}
\widehat{sl}(N) = \oplus_{i} {\cal{G}}_i, \; \qquad
\lbrack Q ,{\cal{G}}_i \rbrack = i  \, {\cal{G}}_i \,. 
\label{gradingstructure}
\end{eqnarray}
This property provides foundation for the zero-curvature approach of
section \ref{section:curvature} used to
determine flows of an integrable hierarchy. %

The  operator $Q$ in (\ref{gradoperator})  induces the following graded subspaces:
\begin{equation}
\begin{split}
 {\cal{G}}_{nN} &=\{ h_1^{(n)}, \cdots ,  h_{N-1}^{(n)}\},  \\
 {\cal{G}}_{nN +1} &=\{ E_{\alpha_1}^{(n)}, \cdots ,  E_{\alpha{_{N-1}}}^{(n)}, E_{-\alpha_1\cdots -\alpha_{N-1}}^{(n+1)}\}, \\
 {\cal{G}}_{nN+2} &=\{ E_{\alpha_1+\alpha_2}^{(n)}, E_{\alpha_2+\alpha_3}^{(n)}, \cdots ,  E_{\alpha_{N-2}+\alpha{_{N-1}}}^{(n)}, E_{-\alpha_1\cdots -\alpha_{N-2}}^{(n+1)},   E_{-\alpha_2\cdots -\alpha_{N-1}}^{(n+1)}\}, \\
\vdots &= \vdots  \\
 {\cal{G}}_{nN +N-1}&=\{ E_{-\alpha_1}^{(n+1)}, \cdots ,  E_{-\alpha{_{N-1}}}^{(n+1)}, E_{\alpha_1 +\cdots+ \alpha_{N-1}}^{(n)}\}. 
 \end{split}
 \label{a4}
\end{equation}
 with each subspace  ${\cal {G}}_m$ of  grade $m\in Z$ containing  generators  of grade $m$.

\section{From periodic dressing chain to Painlev\'e equations}
\label{section:fromj2f}
For completeness and notational consistency we will provide here a number of useful identities to 
prove  that the periodic dressing equation \eqref{mape}
can be rewritten as higher Painlev\'e equations 
\eqref{solut22} and \eqref{evenPeq} for $N$ odd and even, respectively.

Under substitution \eqref{gidef} the dressing equation \eqref{mape} can
be rewritten as 
\begin{equation}
f_{i\, z} =  f_i \big[  f_{i-1} +{\ldots}  - (-1)^k
 f_{i-k} + (-1)^k j_{i-k}
 - f_{i+1} +{\ldots} +(-1)^{k}
 f_{i+k}+ (-1)^{k+1} j_{i+k+1}  \big]  +  \alpha_i  ,    
\label{solut}
\end{equation}
after replacing $j_{i-1}=f_{i-2}-j_{i-2} $ and
$j_{i+2}=f_{i+2}-j_{i+3}$ etc. 

For $N$ odd using the periodicity condition
\eqref{period} we conclude that 
\begin{equation}
j_{i-k}- j_{i+k+1} =0 \;\; \mbox{for} \;\; 2k+1=N \, ,
\label{periodN}
\end{equation}
and therefore equation \eqref{solut} becomes for $2k+1=N$ :
\begin{equation}
f_{i\, z}  = - f_i \lbrack  \sum_{m=1}^{k} (-1)^m f_{i+m} - \sum_{m=1}^{k} (-1)^m
 f_{i-m} \rbrack
 +\alpha_i    \, .
\label{solut1}
\end{equation}
Due to periodicity, $f_{i-m}=f_{i-m+N}$ and we can
rewrite \eqref{solut1} in a standard from : 
\begin{equation}
f_{i\, z} =   f_i \big[ 
\sum_{r=1}^{k}  f_{i+2r-1}  - \sum_{r=1}^{k}  f_{i+2r}  \rbrack  +
\alpha_i \, .
\label{solut2}
\end{equation}

{}From condition \eqref{newcondj} we derive
\begin{equation}
\sum_{i=1}^N f_i = 2 \sum_{i=1}^N j_i=   \frac{N}{2} z \, .
\label{newcondgj}
\end{equation}
Thus it follows that the dressing chain equation  \eqref{mape} gives  $A^{(1)}_{2k}=A^{(1)}_{N-1}$ 
Painlev\'e equations for $N$ odd with $\sum_{i=1}^N f_i =Nz/2$
and $\sum_{i=1}^N \alpha_i =N/2$ for $\alpha_i = \beta_i+ \frac12$.

For $N$ even the condition \eqref{newcondj}
can now be cast as
\begin{equation}
\sum_{r=1}^{n}f_{i+2r}=   \sum_{r=1}^{n}f_{i+2r+1} =\frac{n}{2}z \, .
\label{condsN2nb}
\end{equation}
We next rewrite the dressing equation \eqref{mape} as:
\begin{equation}\label{eom}
f_{i\, z} =f_i \left( 2j_{i+1}-f_i \right) +\alpha_i
, \qquad
i=1,{\ldots} ,N \, .
\end{equation}

To eliminate $j_{i+1}$ from the above equation we sum the derivatives as follows:
\begin{equation}
\sum_{r=1}^{n} f_{i+2r\, z} =\frac{n}{2}=\sum_{r=1}^{n}
2 f_{i+2r}j_{i+2r+1}-\sum_{r=1}^{n} f_{i+2r}^2+\sum_{r=1}^{n}\alpha_{i+2r} \, .
\end{equation}
The squared terms can be eliminated due to relation
\[f_{i}=\sum_{s=1}^{n}f_{i+2s\pm 1}-\sum_{s=1}^{n-1}f_{i+2s}\]
leaving :
\begin{equation}
\sum_{r=1}^{n}\left(\frac12 -\alpha_{i+2r}\right)=
2\sum_{r=1}^{n}f_{i+2r}j_{i+2r+1}-\sum_{r=1}^{n} f_{i+2r}\left(\sum_{s=1}^{n}f_{i+2r+2s\pm 1}-
\sum_{s=1}^{n-1}f_{i+2r+2s}\right) \, .
\end{equation}
To rewrite terms explicitly depending on $j_{i+2r+1}$ we 
use the formula :
\begin{equation}
j_{i+1+2r}-j_{i+1}=\sum_{s=r}^{n-1}f_{i+2s+1}-\sum_{s=r}^{n-1}f_{i+2s+2} \, , 
\end{equation}
that follows from expansion $j_{i+1+2r}=f_{i+2r+1}-f_{i+2r+2}...$
carried until we reach $+f_{i+2n-1}-f_{i+2n}$.

In this way we obtain :
\begin{align}
nz j_{i+1}=&\frac{f_i}{2}(nz)
-2\sum_{r=1}^{n-1} f_{i+2r}\left(\sum_{s=r}^{n-1}f_{i+2s+1}\right)
+2\sum_{r=1}^{n-1} f_{i+2r}\left(\sum_{s=r}^{n-1}f_{i+2s+2}\right) 
\nonumber\\
&+\sum_{r=1}^{n-1}f_{i+2r}
\left(\sum_{s=1}^{n}f_{i+2r+2s\pm 1}\right)
-\sum_{r=1}^{n}f_{i+2r}\left(\sum_{s=1}^{n-1}f_{i+2r+2s}\right)
+\sum_{r=1}^{n}\left(\frac12 -\alpha_{i+2r}\right) \,.\label{4summations}
\end{align}
One can show that the second and the fourth summation terms on the
right hand side of \eqref{4summations}
cancel each other while the difference of the first and the third
summation terms on the
right hand side of \eqref{4summations}
combine to yield :
\begin{subequations}
\begin{align}
&\left(\sum_{r=1}^{n-1}\sum_{s=1}^{r}f_{i+2s-1}f_{i+2r}-\sum_{r=1}^{n-1}f_{i+2r}\left(\sum_{s=r}^{n-1}f_{i+2s+1}\right)\right)\\
&=\left(\sum_{1\leq s\leq r\leq n-1}f_{i+2s-1}f_{i+2r}-\sum_{1\leq r\leq s\leq n-1}f_{i+2r}f_{i+2s+1}\right) \, .
\end{align}
\end{subequations}

To summarize we obtain an expression for $j_{i+1}, i=1, {\ldots} , N$:
\begin{equation}
j_{i+1}=\frac{1}{nz}\left(\frac{f_i}{2}(nz)+
\left(\sum_{1\leq r\leq s\leq n-1}f_{i+2r-1}f_{i+2s}-
\sum_{1\leq r\leq s\leq n-1}f_{i+2r}f_{i+2s+1}\right)+\sum_{r=1}^{n}\left(\frac12 -\alpha_{i+2r}\right)\right) \, .
\end{equation}

Plugging it back into \eqref{eom} yields:
\begin{align}
nzf_{i\, z}=&f_i \left(  \sum_{1\leq r\leq s\leq n-1}f_{i+2r-1}f_{i+2s}
-\sum_{1\leq r\leq s\leq n-1}f_{i+2r}f_{i+2s+1}+\sum_{r=1}^{n}\left(1 -2\alpha_{i+2r}\right) 
\right) \nonumber\\
&+nz \alpha_i, \qquad
i=1,{\ldots} ,N \, ,
\end{align}
reproducing $A^{(1)}_{2n-1}=A^{(1)}_{N-1}$ 
invariant Painlev\'e equations for even $N=2n$  with 
conditions 
\eqref{condsN2nb}.

\vspace{5mm}
{\bf Acknowledgments}
JFG and AHZ thank CNPq and FAPESP for financial support. VCCA 
thanks S\~ao Paulo Research Foundation (FAPESP) for financial support 
by grants 2016/22122-9 and 2019/03092-0.


\begin{thebibliography}{99}
%
\bibitem{adler93}
 Adler V E  1993 %
Functional Analysis and Its Applications, {\bf 27}:2, 141--143 
%
\bibitem{victorsthesis}
Alves V C C 2021  \textit{On Hybrid Painlev\'e Equations}, Doctoral dissertation (S\~ao Paulo, Brazil-UNESP)
\bibitem{coalescence}
Alves V C C, Aratyn H, Gomes J F and Zimerman A H 2020
\textit{J. Phys. A: Math. Theor.} \textbf{53} 445202
\bibitem{p3-p5s} Alves V C C, Aratyn H, Gomes J F and Zimerman A H 2019  \textit{J. Phys. A: Math. Theor.} \textbf{52} 065203, and 2019
\textit{J. Phys.: Conf. Ser.} \textbf{1194} 012002
\bibitem{p3-p5}
Aratyn H, Gomes J F, Ruy D V and Zimerman A H 2016
\textit{J. Phys. A: Math. Theor.} \textbf{49} 045201
\bibitem{higherpainleve}
Aratyn H, Gomes J F and Zimerman A H 2011 \textit{ J. Phys. A: Math. Theor.} \textbf{44}
235202
\bibitem{anpz} Aratyn H, Nissimov E, Pacheva S and Zimerman A H 1995
\textit{Int. J. Mod. Phys. A} \textbf{10} 2537, arXiv:hep-th/9407112
\bibitem{Aratyn:1993zi}
Aratyn H, Ferreira L A, Gomes J F and Zimerman A H 1993
%
%
\textit{Phys. Lett.}  {\bf  B 316} 85,
arXiv:hep-th/9307147
%
%
%
%
\bibitem{globo}
Carvalho Ferreira J M, Gomes J F,  Lobo G V and   Zimerman A H 2021
%
{\it J. Phys. A: Math. Theor. }{\bf 54},  doi.org/10.1088/1751-8121/abd8b2,   
arXiv:2010.03631 [nlin.SI]
 	\bibitem{cosgrove}
 	Cosgrove C M 2006 \textit{Studies in Applied Mathematics} \textbf{116} 321-413
\bibitem{flaschka}
Flaschka H and Newell A C 1980 \textit{Comm. Math. Phys.} \textbf{76} 65
\bibitem{jfg2015}
Gomes J F, Retore A L and Zimerman A H 2015  
%
J. Phys. A: Math. Theor. \textbf{48}  405203, arXiv:1505.01024
\bibitem{jfg2016}
Gomes  J F, Retore A L and Zimerman A H 2016
%
J. Phys. A: Math. Theor. \textbf{49} 504003,  arXiv:1610.02303 
\bibitem{fordy}
Fordy A and Gibbons J 1980 
%
\textit{Commun. Math. Phys.} \textbf{77} 21-30
\bibitem{ince}
Ince E L 1956 \textit{Ordinary Differential Equations} (New York: Dover)
\bibitem{alred}
Noumi M and Yamada Y 1998 \textit{Commun. Math. Phys.} \textbf{199} 281-195,
	 %
%
\textit{Funkcial.Ekvac.} \textbf{41} 483- 503 arXiv:math/9808003
%
%
%
%
%
\bibitem{schiff93}
Schiff J 1994 Nonlinearity, {\bf 7}, 305-312 
%
\bibitem{SHC06}
Sen A,  Hone A N W and  Clarkson P A 2006
%
Studies in Applied Mathematics {\bf 117} 299-319
\bibitem{takasaki03}
Takasaki K 2003  %
%
Commun. in Mathematical Physics, {\bf 241}, 111-142  
\bibitem{Tsuda04}
Tsuda T 2005 Advances in Mathematics {\bf 197}  587- 606
%
%
%
%
\bibitem{Veselov-Shabat}
Veselov A P and Shabat A B 1993 \textit{Funct. Anal. Appl.} \textbf{27} 81--96
\bibitem{WH03}
Willox R and Hietarinta J 2003
%
\textit{J. Phys. A: Math. Gen.} \textbf{36}, 10615--10635
\end{thebibliography}
\end{document}